\documentclass[twocolumn]{aastex701}

\usepackage{amsmath}
\usepackage{booktabs}
\usepackage{microtype}

\graphicspath{{./}}
\newcommand{\pimen}{\ensuremath{\pi}\ Mensae}

\shorttitle{The Orbit of \pimen\ d}
\shortauthors{Jiang}

\begin{document}

\title{An Outer Giant Excites and Constrains the Orbit of \pimen\ d}

\correspondingauthor{Qunfeng Jiang}
\author[0009-0008-0600-3743]{Qunfeng Jiang}
\affiliation{Independent Researcher, Shanghai, China}
\email[show]{qfjiang01@gmail.com}

\begin{abstract}
The nearby \pimen\ system contains the eccentric giant planet b on a 5.7 yr orbit, the 6.27 day transiting sub-Neptune c, and a candidate planet d on a 123 day orbit between them.  Published estimates place the eccentricity of d between approximately 0.2 and 0.4.  We reanalyze 772 published velocities with three-Keplerian fits and explore three-dimensional b--d dynamics with $N$-body integrations.  The high-eccentricity solution persists when the phase or apsidal prior is widened; exact importance reweighting to a population-informed Beta eccentricity prior gives $e_d=0.378^{+0.050}_{-0.049}$.  In 20 kyr integrations, 84.5\% of configurations near the prograde and retrograde coplanar endpoints survive, compared with 21.3\% across $60^\circ\leq I_{bd}<120^\circ$; stellar collisions dominate the failures.  Secular forcing by b moves 53 of 62 one-Myr survivors initialized with $e_d\leq0.1$ into $0.3\leq e_d\leq0.5$.  Representative b+d configurations remain viable for 100 Myr, whereas controls with c and general relativity retain fewer systems.  The outer giant can therefore excite the observed eccentricity of d while constraining its viable orbit toward prograde or retrograde near-coplanar configurations.
\end{abstract}


\section{Introduction}\label{sec:introduction}

Radial-velocity observations of the nearby G-type star \pimen\ revealed a massive planet on a highly eccentric 5.7 yr orbit, \pimen\ b \citep{Jones2002}.  Nearly two decades later, the Transiting Exoplanet Survey Satellite (TESS; \citealt{RickerEtAl2015}) discovered \pimen\ c, a 6.27 day transiting sub-Neptune and the first planet reported by the mission \citep{HuangEtAl2018,GandolfiEtAl2018}.  Astrometry and radial velocity (RV) measurements give b a true mass near the planetary--brown-dwarf boundary and show a substantial misalignment between the orbital planes of b and c \citep{DamassoEtAl2020,XuanWyatt2020,DeRosaEtAl2020}.  The system is therefore a compact example of the secular coupling that outer giant planets can impose on inner planetary systems \citep{BryanEtAl2019}.

The candidate planet d occupies the wide gap between c and b.  \citet{HatzesEtAl2022} measured $P_d=124.64^{+0.48}_{-0.52}$ days, $K_d=1.68\pm0.17\ \mathrm{m\,s^{-1}}$, and $e_d=0.220\pm0.079$.  An independent analysis found $e_d=0.16\pm0.15$ from a different data selection \citep{LaliotisEtAl2023}.  \citet{LarsenEtAl2026} subsequently reported $P_d=123.1\pm0.2$ days and $e_d=0.41\pm0.05$.  Their eccentricity estimate used an informative conjunction-time prior and restricted the argument of pericenter to $180^\circ$--$360^\circ$.  These measurements establish a plausible high-eccentricity solution amid appreciable posterior and analysis dependence.

The orbit of d is dynamically coupled to the massive and eccentric planet b.  Large mutual inclinations can drive Kozai--Lidov eccentricity--inclination exchange \citep{Kozai1962,Lidov1962}, while the eccentric outer orbit introduces octupole-order modulation and chaotic trajectories \citep{LithwickNaoz2011,KatzEtAl2011,NaozEtAl2013,Naoz2016}.  General-relativistic apsidal precession competes with the forcing from b for planet c and limits the eccentricity growth found in published b+c integrations \citep{XuanWyatt2020,DeRosaEtAl2020}.  For d, the wider orbit makes relativistic precession much slower.

The small RV semiamplitude of d also makes its eccentricity sensitive to sampling, phase coverage, noise assumptions, and the eccentricity prior \citep{ShenTurner2008,HaraEtAl2019}.  Population-informed priors reduce the usual positive eccentricity bias for weak signals \citep{Kipping2013,StevensonEtAl2025}, although a population distribution supplies limited information about an individual orbit.  A dynamical analysis should retain the joint orbital correlations and explore the unknown three-dimensional orientation.

This paper combines an RV prior-sensitivity analysis with a hierarchical dynamical analysis.  Section~\ref{sec:methods} describes the published velocities, orbital fits, initial conditions, and integrations.  Section~\ref{sec:results} presents the RV constraints, inclination-dependent stability, full-system controls, secular excitation from low initial eccentricity, relativistic precession scales, and long-term representative solutions.  Section~\ref{sec:discussion} interprets the viable architectures and compares them with previous dynamical limits.  Section~\ref{sec:summary} summarizes the main constraints.

\section{Data and Methods}\label{sec:methods}

\subsection{Radial-velocity data and orbital fits}\label{sec:rv}

We assembled 772 machine-readable RV measurements from the original electronic tables.  The sample contains 42 measurements from the University College London Echelle Spectrograph (UCLES) in Table~A.1 of \citet{GandolfiEtAl2018}, 402 from the High Accuracy Radial velocity Planet Searcher (HARPS) in the electronic RV table of \citet{HatzesEtAl2022}, and 275 from the Echelle SPectrograph for Rocky Exoplanets and Stable Spectroscopic Observations (ESPRESSO) plus 53 from CORALIE in Tables~B.1 and B.2 of \citet{DamassoEtAl2020}.  The three published CORALIE epochs contain 10, 11, and 32 velocities.  \citet[][their Table~6]{LarsenEtAl2026} list 60 CORALIE measurements and a total of 779; the seven-row difference lies entirely in the reported CORALIE count because the UCLES, HARPS, and ESPRESSO counts agree.  Our analysis uses the 772 velocities available in the cited electronic tables.

The RV model is the sum of three independent Keplerian signals for b, c, and d.  Each of the six instrumental series has its own additive velocity offset and white-jitter term.  At every likelihood evaluation, the offset is the inverse-variance-weighted mean residual for that series.  The likelihood is
\begin{equation}
\begin{split}
\ln \mathcal{L}=-\frac{1}{2}\sum_i\bigg[
&\frac{(v_i-m_i-\widehat{\gamma}_{k(i)})^2}{\sigma_i^2+s_{k(i)}^2}\\
&+\ln\{2\pi(\sigma_i^2+s_{k(i)}^2)\}\bigg],
\end{split}
\end{equation}
where $m_i$ is the three-planet Keplerian model, $\widehat{\gamma}_{k(i)}$ is the fitted offset, and $s_{k(i)}$ is the jitter for the relevant series.  The jitter priors are log-uniform from 0.01 to $100\ \mathrm{m\,s^{-1}}$.  We use \textsc{RadVel} \citep{FultonEtAl2018} for Keplerian evaluation and for converting between conjunction and pericenter times.  Posterior sampling uses the static \textsc{dynesty} nested sampler with 1000 live points, multi-ellipsoid bounds, random-walk proposals, and a stopping threshold $\Delta\ln Z=0.5$ \citep{Speagle2020}.

The reference fit follows the conjunction-time and apsidal priors used by \citet{LarsenEtAl2026}.  Two independently sampled sensitivity fits change one prior at a time by widening either $\omega_d$ or the conjunction-time prior.  To test whether the relatively high inferred eccentricity is sensitive to the uniform eccentricity prior, we also use a population-informed prior that favors the lower eccentricities commonly found among RV planets.  For this matched eccentricity-prior test, we multiply the original reference-posterior weights by
\begin{equation}
\frac{\pi_{\rm Beta}(e_d)}{\pi_{\rm U}(e_d)},
\end{equation}
where $\pi_{\rm U}$ is uniform on $0<e_d<0.8$ and $\pi_{\rm Beta}$ is the $\mathrm{Beta}(0.867,3.03)$ distribution of \citet{Kipping2013}, normalized over the same interval.  This importance reweighting holds the data, likelihood, parameterization, support, and all other priors fixed.

\subsection{Initial conditions and numerical model}\label{sec:initial}

For each dynamical realization, the orbital elements of b and d are drawn jointly from high-eccentricity RV posterior samples, preserving their fitted correlations.  The posterior-based dynamical screen uses two equally sized input ensembles: a uniform-$e_d$ ensemble with a median eccentricity of approximately 0.412 and a Beta-$e_d$ ensemble with a median of approximately 0.377.  The latter also adopts wider conjunction-time and apsidal priors.  Each ensemble contributes 1000 realizations to the 20 kyr screen.  The parameters used to initialize the integrations are summarized in Table~\ref{tab:initial}.  We draw the inclination of b from a split-normal distribution centered on $45.8^\circ$, with upper and lower scales of $1.4^\circ$ and $1.1^\circ$, consistent with the astrometric orbit \citep{DamassoEtAl2020,DeRosaEtAl2020}.  A rotation sets $\Omega_b=0$.  The pole of d is isotropic: $\cos i_d$ is uniform on $[-1,1]$ and $\Omega_d$ is uniform on $[0,2\pi)$.  The b--d mutual inclination is
\begin{equation}
\cos I_{bd}=\cos i_b\cos i_d+\sin i_b\sin i_d\cos(\Omega_b-\Omega_d).
\end{equation}
We derive the true masses of b and d by numerically solving the exact RV mass function using the corresponding sampled $P$, $e$, $K$, and $i$, together with the adopted stellar mass.  Orbital phases are evaluated at BJD 2459000.0 from the fitted conjunction times; the stellar argument of pericenter is shifted by $180^\circ$ when constructing the planet orbit.  The elements are inserted relative to the star and transformed to the barycentric frame before integration; $R_\star$ sets the stellar-collision boundary.

\begin{deluxetable*}{ll}
\tablenum{1}
\tablecaption{Initial Conditions for the Dynamical Experiments\label{tab:initial}}
\tabletypesize{\scriptsize}
\tablehead{\colhead{Parameter} & \colhead{Value or distribution}}
\startdata
\multicolumn{2}{c}{Star} \\
$M_\star$ & $1.07\,M_\odot$\tablenotemark{a} \\
$R_\star$ & $1.19\,R_\odot=0.00553$ au\tablenotemark{a} \\
\tableline
\multicolumn{2}{c}{b} \\
$P_b$ & Joint RV-posterior draw \\
$e_b$ & Joint RV-posterior draw \\
$K_b$ & Joint RV-posterior draw \\
$\omega_b$ & Joint RV-posterior draw \\
$t_{{\rm conj},b}$ & Joint RV-posterior draw \\
$i_b$ & $45.8^{+1.4}_{-1.1}$ deg\tablenotemark{b} \\
$\Omega_b$ & $0^\circ$ \\
$M_b$ & Derived from $P_b,e_b,K_b,i_b,$ and $M_\star$ \\
\tableline
\multicolumn{2}{c}{c} \\
$P_c$ & $6.267852$ days\tablenotemark{a} \\
$M_c$ & $3.63\,M_\oplus$\tablenotemark{a} \\
$e_c$ & 0\tablenotemark{a} \\
$i_c$ & $87.05^\circ$\tablenotemark{a} \\
$t_{{\rm conj},c}$ & 2458519.8068 BJD\tablenotemark{a} \\
$\Omega_c$ & $\mathrm{U}(0^\circ,360^\circ)$ \\
\tableline
\multicolumn{2}{c}{d} \\
$P_d$ & Joint RV-posterior draw \\
$e_d$ & Joint RV-posterior draw \\
$K_d$ & Joint RV-posterior draw \\
$\omega_d$ & Joint RV-posterior draw \\
$t_{{\rm conj},d}$ & Joint RV-posterior draw \\
$\cos i_d$ & $\mathrm{U}(-1,1)$ \\
$\Omega_d$ & $\mathrm{U}(0^\circ,360^\circ)$ \\
$M_d$ & Derived from $P_d,e_d,K_d,i_d,$ and $M_\star$
\enddata
\tablenotetext{a}{Adopted from \citet{HatzesEtAl2022}.}
\tablenotetext{b}{Astrometric inclination from \citet{DamassoEtAl2020} and \citet{DeRosaEtAl2020}.}
\tablecomments{The RV parameters for b and d are drawn jointly to retain their posterior correlations.  The inclination of b is sampled from a split-normal distribution using the quoted asymmetric scales.  Two independent values of $\Omega_c$ are drawn for each selected b+d configuration.}
\end{deluxetable*}

The reference model contains planets b and d and uses REBOUND with the adaptive IAS15 integrator and accuracy parameter $\epsilon=10^{-9}$ \citep{ReinLiu2012,ReinSpiegel2015}.  The full-system controls add planet c with the parameters in Table~\ref{tab:initial}.  General relativity (GR) is included with REBOUNDx \citep{TamayoEtAl2020}: the paired b+d+GR controls use the \texttt{gr} module with IAS15, whereas the b+c+d+GR controls use the conservative \texttt{gr\_potential} module with the hybrid MERCURIUS integrator and a base step of $P_c/30$ \citep{ReinEtAl2019}.

An integration ends when any included planet---b, c, or d---collides with the star, two planets collide, the b--d separation falls within their mutual Hill radius evaluated at b's initial pericenter, an osculating orbit becomes unbound, or a body crosses 100 au.  A trajectory that reaches the requested duration is classified as a survivor.  Tides are neglected because illustrative equilibrium-tide estimates give eccentricity-damping timescales of $\sim10^9$ yr for low-eccentricity c and $\sim10^{15}$ yr for d near its current orbit, much longer than the respective integration durations \citep{Hut1981,LeconteEtAl2010}.  Stellar collisions terminate the integrations and are classified as unstable, while survivors reaching pericenters of a few stellar radii are retained as point-mass survivors potentially subject to tidal modification \citep{FabryckyTremaine2007,LiuEtAl2015}.

\subsection{Dynamical experiments}\label{sec:experiments}

First, we use the staged integrations summarized in Table~\ref{tab:posteriorflow} to identify present-day b+d configurations that remain viable over progressively longer intervals.  Next, we test whether the eccentric giant b can excite an initially near-circular d and vary $e_b$ to distinguish the effects of high mutual inclination from those of b's eccentricity.  Finally, we use paired models to assess the effects of c and GR and recompute two long-term trajectories at tighter IAS15 accuracy.

\begin{deluxetable}{lrrr}
\tablenum{2}
\tablecaption{Sample Sizes and Survival in the Posterior-based b+d Integrations\label{tab:posteriorflow}}
\tabletypesize{\scriptsize}
\tablehead{\colhead{Stage} & \colhead{Integrations} & \colhead{Duration} & \colhead{Survivors}}
\startdata
Initial screen & 2000 & 20 kyr & 800 \\
Random extension & 200 & 1 Myr & 81 \\
Inclination-binned extension & 39 & 10 Myr & 12 \\
Extension of all 10 Myr survivors & 12 & 50 Myr & 9 \\
Representative extension & 6 & 100 Myr & 5
\enddata
\tablecomments{Random extension: 100 randomly selected 20 kyr survivors from each RV input ensemble.  Inclination-binned extension: within each ensemble, the one-Myr survivors were divided into four mutual-inclination bins, and up to five systems were randomly selected from each nonempty bin.}
\end{deluxetable}

\textit{Posterior geometry and long-term integrations.}  To determine which present-day configurations are allowed and how survival depends on three-dimensional orientation, we first integrated all 2000 posterior b+d configurations for 20 kyr; 800 survived.  From those survivors, 100 cases from each RV input distribution were selected by simple random sampling without replacement and extended to 1 Myr, leaving 81 survivors.  To retain coverage across the full mutual-inclination range, we treated the two RV input distributions separately and divided the one-Myr survivors in each into four bins of initial mutual inclination: $0^\circ$--$30^\circ$, $30^\circ$--$90^\circ$, $90^\circ$--$150^\circ$, and $150^\circ$--$180^\circ$.  We randomly selected up to five systems from each nonempty bin, yielding 39 systems for the 10 Myr integrations; 12 survived.  All 12 were continued to 50 Myr, and nine survived.  For the 100 Myr integrations, the numerically qualified 50 Myr survivors were grouped into the same four inclination bins.  We randomly selected one system from each nonempty bin, then selected additional systems at random from the remaining eligible pool to obtain six configurations.

\textit{Excitation from low eccentricity.}  To test whether secular forcing by the eccentric giant planet b can excite an initially near-circular orbit of d into the observationally relevant eccentricity range, we also performed a controlled low-eccentricity experiment.  We set
\begin{equation}
e_{d,0}=0,\ 0.02,\ 0.05,\ 0.10
\end{equation}
At each value, we generated 64 configurations from the distributions in Table~\ref{tab:loweinit}, for 256 in total.  A scrambled Sobol sequence was used to distribute the configurations more evenly across the allowed parameter space.  All 256 systems were integrated for 20 kyr.  Up to 32 survivors per $e_{d,0}$ group were then randomly selected without replacement, yielding 110 systems for extension to 1 Myr.  Sixty-two survived to 1 Myr, and 53 of those reached $0.3\leq e_d\leq0.5$ at least once.

\begin{deluxetable}{ll}
\tablenum{3}
\tablecaption{Initial Conditions for the Low-$e_d$ Experiment\label{tab:loweinit}}
\tabletypesize{\scriptsize}
\tablehead{\colhead{Parameter} & \colhead{Value or distribution}}
\startdata
\multicolumn{2}{c}{b} \\
$P_b$ & $2088.33$ days\tablenotemark{a} \\
$e_b$ & 0.6396\tablenotemark{a} \\
$M_b$ & $12.6\,M_{\rm Jup}$\tablenotemark{a} \\
$i_b$ & $0^\circ$ \\
$\Omega_b$ & $0^\circ$ \\
$\omega_{b,\star}$ & $331.03^\circ$\tablenotemark{a} \\
$t_{{\rm conj},b}$ & 2456540.34 BJD\tablenotemark{a} \\
\tableline
\multicolumn{2}{c}{d} \\
$P_d$ & $124.64$ days\tablenotemark{a} \\
$M_d$ & $13.38\,M_\oplus$\tablenotemark{a} \\
$e_{d,0}$ & $0,\ 0.02,\ 0.05,\ 0.10$ \\
$\cos i_d$ & $\mathrm{U}(-1,1)$ \\
$\Omega_d$ & $\mathrm{U}(0^\circ,360^\circ)$ \\
$\omega_{d,\star}$ & $\mathrm{U}(0^\circ,360^\circ)$ \\
$\mathcal{M}_d$ & $\mathrm{U}(0^\circ,360^\circ)$
\enddata
\tablenotetext{a}{Adopted from \citet{HatzesEtAl2022}.}
\tablecomments{At each $e_{d,0}$ value, 64 configurations are generated from the distributions shown.}
\end{deluxetable}

\textit{Dependence on the eccentricity of b.}  To separate the effect of high mutual inclination from that of b's measured eccentricity, we used the outcomes of the preceding integrations to select six representative cases: a stable prograde case, a stable retrograde case, a strongly excited survivor, an early failure, a delayed failure, and a near-polar failure.  For each case, all three $e_b$ controls were restarted from that case's original initial condition, with the earlier outcome serving only to assign the representative category.  For each selected configuration we used $e_b$ at its RV-drawn value ($\simeq0.64$), 0.2, and 0, giving 18 one-Myr b+d integrations.  Only $e_b$ was changed; the initial mean anomaly of b and all other initial parameters were retained.

\textit{Planet c and relativistic precession.}  The effect of c and GR was tested with 25 selected b+d configurations: 10 that survived 10 Myr, 10 that failed between 1 and 10 Myr, and five that failed within 20 kyr.  Each configuration retained the same original b+d initial condition in four paired one-Myr models: b+d, b+d+GR, and b+c+d+GR at each of its two independently drawn $\Omega_c$ values.  For each b+d configuration, $\Omega_1$ and $\Omega_2$ denote its own two independently sampled values of $\Omega_c$, so the pair varies from one configuration to another.  The four 25-member sets retained 20, 15, 6, and 7 survivors, respectively.

\textit{Numerical cross-checks.}  Finally, one strongly excited survivor and the delayed-failure case from the 100 Myr sample were recomputed with the IAS15 accuracy parameter $\epsilon$ reduced from $10^{-9}$ to $10^{-11}$, imposing a stricter criterion for adaptive timestep selection.

\section{Results}\label{sec:results}

\subsection{Radial-velocity constraints}\label{sec:rvresults}

\begin{figure*}[t!]
\centering
\includegraphics[width=\textwidth]{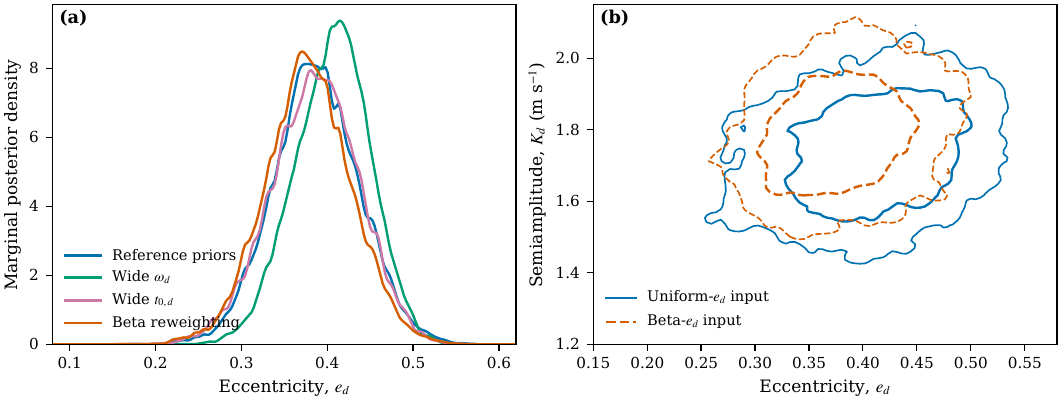}
\caption{RV constraints on the eccentricity of \pimen\ d.  Panel (a) compares the reference posterior, two independently sampled one-prior sensitivity fits, and the exact prior-ratio reweighting of the reference samples to the truncated Beta eccentricity prior.  Panel (b) shows the 68\% and 95\% contours in $(e_d,K_d)$ for the two RV distributions from which the dynamical initial conditions were drawn.\label{fig:rvpost}}
\end{figure*}

\begin{deluxetable}{lccc}
\tablenum{4}
\tablecaption{RV Prior-sensitivity Analysis\label{tab:rvpriors}}
\tablehead{\colhead{Fit} & \colhead{$P_d$ (day)} & \colhead{$K_d$ (m s$^{-1}$)} & \colhead{$e_d$}}
\startdata
Reference priors & $123.209^{+0.471}_{-0.200}$ & $1.735^{+0.115}_{-0.114}$ & $0.387^{+0.051}_{-0.049}$ \\
Wide $\omega_d$ & $123.498^{+0.375}_{-0.378}$ & $1.678^{+0.103}_{-0.103}$ & $0.408^{+0.041}_{-0.046}$ \\
Wide $t_{0,d}$ & $122.971^{+0.207}_{-0.236}$ & $1.781^{+0.119}_{-0.121}$ & $0.386^{+0.050}_{-0.052}$ \\
Beta reweighting & $123.223^{+0.494}_{-0.209}$ & $1.731^{+0.115}_{-0.113}$ & $0.378^{+0.050}_{-0.049}$
\enddata
\tablecomments{Values are medians and 16th--84th percentiles for the 772 published velocities.  The wide-$\omega_d$ and wide-$t_{0,d}$ rows are independent fits that each change one prior.  The last row instead reweights the original reference samples from a uniform prior on $0<e_d<0.8$ to a $\mathrm{Beta}(0.867,3.03)$ prior truncated to the same support.}
\end{deluxetable}

The reference fit yields $e_d=0.387^{+0.051}_{-0.049}$.  Widening the apsidal prior gives $e_d=0.408^{+0.041}_{-0.046}$, while widening the conjunction-time prior gives $e_d=0.386^{+0.050}_{-0.052}$.  Reweighting the reference samples to the population-informed Beta eccentricity prior, normalized over the same $0<e_d<0.8$ interval, gives $e_d=0.378^{+0.050}_{-0.049}$.  Table~\ref{tab:rvpriors} summarizes these constraints, and Figure~\ref{fig:rvpost}a compares the corresponding marginal eccentricity distributions.

The reference, wide-$\omega_d$, and wide-$t_{0,d}$ fits have median eccentricities between 0.386 and 0.408.  For the solution represented by the reference samples, and with the eccentricity support and all other priors held fixed, Beta reweighting lowers the inferred eccentricity only modestly, leaving it close to 0.4.  The high-eccentricity solution therefore persists under the phase, apsidal, and eccentricity-prior variations examined here.

\subsection{Inclination-dependent survival}\label{sec:orientation}

The posterior-based survival analysis summarized in Table~\ref{tab:posteriorflow} proceeds in two stages.  We first integrate 2000 Newtonian b+d realizations for 20 kyr, with 1000 from each of the uniform-$e_d$ and Beta-$e_d$ input ensembles described in Section~\ref{sec:initial} and shown in Figure~\ref{fig:rvpost}b.  We then randomly select 100 survivors from each ensemble and extend these same trajectories to a total duration of 1 Myr.

\begin{figure*}[t!]
\centering
\includegraphics[width=0.86\textwidth]{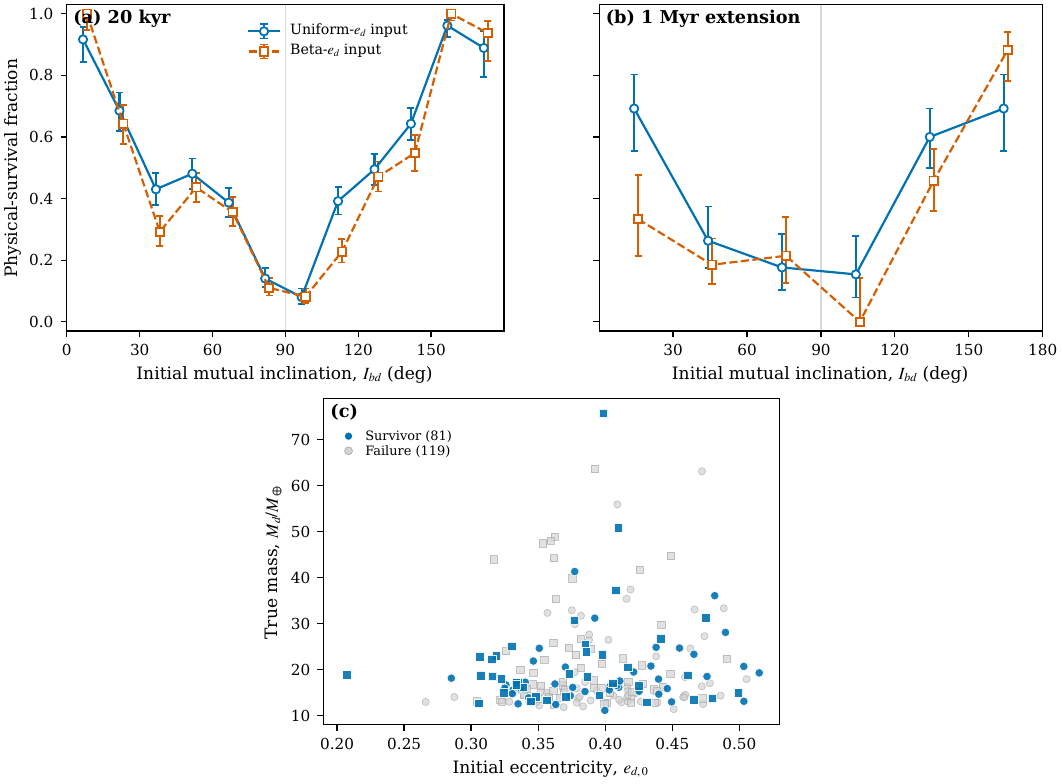}
\caption{Outcomes of the posterior-based b+d integrations.  Panel (a) shows the physical-survival fraction in the 20 kyr screen as a function of initial b--d mutual inclination, in $15^\circ$ bins.  Panel (b) shows the one-Myr outcomes of the randomly selected 20 kyr survivors in $30^\circ$ bins.  Error bars in panels (a) and (b) are 68\% Wilson intervals, and lines guide the eye.  Panel (c) shows the initial eccentricity and true mass of d for the 200 extended systems.  One-Myr survivors are highlighted, while failures are shown in gray; circles denote the uniform-$e_d$ input and squares denote the Beta-$e_d$ input.\label{fig:ushape}}
\end{figure*}

Figure~\ref{fig:ushape}a shows the inclination-dependent survival in the initial screen.  Of the 2000 systems, 800 survive to 20 kyr.  Among the 1200 failures, 1198 are collisions between d and the star, and the remaining two are close encounters.  Combining the two input ensembles, 84.5\% of configurations with $I_{bd}<30^\circ$ or $I_{bd}\geq150^\circ$ survive, compared with 21.3\% across $60^\circ\leq I_{bd}<120^\circ$.  The depletion is deepest near a perpendicular b--d geometry, where large-amplitude Kozai--Lidov oscillations can drive d to extreme eccentricities and small pericenters, leading to stellar collisions (see Section~\ref{sec:becc}).

Figure~\ref{fig:ushape}b shows the outcomes of the random extensions.  Of the 200 selected 20 kyr survivors, 81 remain at 1 Myr.  Within this extension sample, the conditional survival fraction is 67.3\% near the two coplanar endpoints and 16.0\% across $60^\circ\leq I_{bd}<120^\circ$, preserving the strong inclination dependence found in the initial screen.  These fractions describe subsequent survival among the selected 20 kyr survivors.

In the $0^\circ\leq I_{bd}<30^\circ$ bin, 9/13 uniform-$e_d$ systems and 4/12 Beta-$e_d$ systems survived to 1 Myr.  The corresponding 20 kyr fractions were similar (59/78 and 54/74).  Because the 1 Myr samples are small and the input ensembles also differ in phase and apsidal priors, the contrast is consistent with finite-sample variation.

Figure~\ref{fig:ushape}c compares the initial eccentricities and true masses of the extended systems, distinguishing the one-Myr survivors from the failures.  The eccentricities are drawn from the respective joint RV posteriors.  One-Myr survivors include initial eccentricities as high as $e_{d,0}=0.515$ and true masses as high as $75.7\,M_\oplus$.  These cases show that favorable b--d geometries can accommodate eccentric or relatively massive realizations of d.

\subsection{Planet c and general relativity}\label{sec:cgr}

For each of the 25 selected b+d configurations, we integrated the four models described in Section~\ref{sec:experiments} for 1 Myr, yielding 100 integrations with matched original b+d initial conditions.  Table~\ref{tab:cgrcontrols} groups the survival counts by the configurations' previous outcomes in the Newtonian b+d model.

Here $\Omega_1$ and $\Omega_2$ are the configuration-specific pair of independently sampled $\Omega_c$ values defined in Section~\ref{sec:experiments}; the pair varies across configurations.

\begin{deluxetable*}{lcccc}
\tablenum{5}
\tablecaption{Survival in the Matched Planet-c and GR Controls\label{tab:cgrcontrols}}
\tabletypesize{\scriptsize}
\tablehead{\colhead{Previous b+d outcome} & \colhead{b+d} & \colhead{b+d+GR} & \colhead{b+c+d+GR ($\Omega_1$)} & \colhead{b+c+d+GR ($\Omega_2$)}}
\startdata
Survived 10 Myr & 10/10 & 8/10 & 5/10 & 5/10 \\
Failed at 1--10 Myr & 10/10 & 7/10 & 1/10 & 2/10 \\
Failed within 20 kyr & 0/5 & 0/5 & 0/5 & 0/5 \\
Total & 20/25 & 15/25 & 6/25 & 7/25
\enddata
\tablecomments{$\Omega_1$ and $\Omega_2$ denote the two independently drawn values of $\Omega_c$ for each configuration.  All entries give the number surviving at 1 Myr.}
\end{deluxetable*}

Adding GR changes the survival outcome in five of the 25 pairs, reducing the number of survivors from 20 to 15.  The median absolute change in $e_{d,\max}$ is only 0.0072.  Additional apsidal precession can shift the secular phase and the timing of extreme pericenter passages, changing finite-time outcomes near the stability boundary even when the typical eccentricity amplitude changes little.

Adding c produces nine and eight additional failures relative to b+d+GR for $\Omega_1$ and $\Omega_2$, respectively.  Planet c is the first planet lost in six and five of these cases.  Among configurations surviving both b+d+GR and the corresponding three-planet model, the median absolute changes in $e_{d,\max}$ are 0.0081 and 0.0028.  Thus, c has a small typical effect on d's eccentricity amplitude among common survivors, while excluding geometries that cannot also preserve the inner planet.

The characteristic precession scales explain why GR acts differently on c and d \citep{FabryckyTremaine2007,Naoz2016}.  The leading relativistic apsidal precession rate is
\begin{equation}
\dot{\omega}_{\rm GR}=\frac{3GM_\star n}{a c_{\rm light}^{2}(1-e^{2})},
\end{equation}
where $n$ is the orbital mean motion and $c_{\rm light}$ is the speed of light.  At the nominal orbital parameters, the GR precession period and the b-driven quadrupole secular timescale are 36.9 and 77.8 kyr for c, compared with 4.38 Myr and 3.96 kyr for d.  GR therefore competes with planetary precession for c, whereas b dominates the secular excitation of d.  Overall, GR modifies the apsidal dynamics and can change phase-sensitive survival outcomes, while c imposes the additional requirement that the viable architecture preserve all three planets.

\subsection{Excitation from low initial eccentricity}\label{sec:lowe}

\begin{figure*}[t!]
\centering
\includegraphics[width=\textwidth]{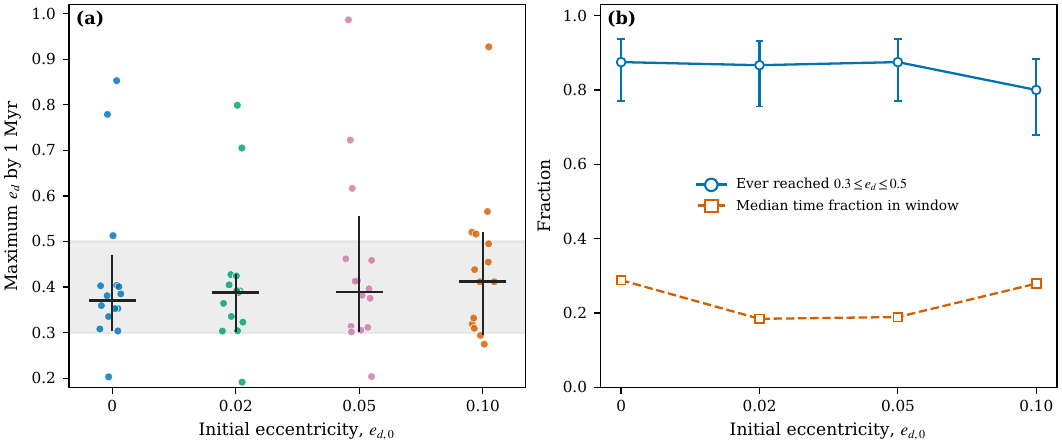}
\caption{Secular excitation in the low-initial-eccentricity experiment.  Panel (a) gives the maximum $e_d$ reached by each one-Myr survivor; vertical bars span the 16th--84th percentiles and horizontal bars mark medians.  The shaded band denotes $0.3\leq e_d\leq0.5$.  Panel (b) compares the fraction of survivors that ever enter this interval with the median fraction of sampled integration time spent there.  Error bars on the blue fractions are 68\% Wilson intervals.  Time fractions are estimated from outputs recorded every 100 yr, and their medians are computed over the one-Myr survivors in each initial-eccentricity group.\label{fig:lowe}}
\end{figure*}

Figure~\ref{fig:lowe} summarizes the excitation of d from low initial eccentricity.  The experiment starts with 256 configurations, comprising 64 at each of $e_{d,0}=0$, 0.02, 0.05, and 0.10, with the remaining parameters drawn as specified in Table~\ref{tab:loweinit}.  After the 20 kyr screen, up to 32 survivors per initial-eccentricity group were randomly selected without replacement, yielding 110 systems for extension to a total duration of 1 Myr.  Of these, 62 survive.

Figure~\ref{fig:lowe}a shows the maximum eccentricities reached by the one-Myr survivors.  Fifty-three of the 62 survivors (85.5\%) enter $0.3\leq e_d\leq0.5$ at least once, including 14 of the 16 survivors initialized on circular orbits.  Figure~\ref{fig:lowe}b compares these entry fractions with the median time fractions spent in the same interval.

For each surviving trajectory, the time fraction is estimated from orbital states recorded at 100 yr intervals:
\begin{equation}
\widehat f_{\rm win}=\frac{N_{\rm win}}{N_{\rm out}},
\end{equation}
where $N_{\rm win}$ is the number of recorded epochs with $0.3\leq e_d\leq0.5$, and $N_{\rm out}$ is the total number of recorded epochs.  Across the 62 survivors, the median estimated time fraction is 22.5\%, with similar medians in the four initial-eccentricity groups.  Secular forcing by b can therefore bring initially near-circular d orbits into the observationally relevant range and maintain them there for an appreciable fraction of their evolution.

\subsection{Dependence on the eccentricity of planet b}\label{sec:becc}

To distinguish inclination-driven excitation from the additional effects of b's eccentricity, we compared 18 one-Myr integrations based on six representative b+d configurations.  Each configuration was integrated with $e_b$ at its RV-drawn value ($\simeq0.64$), 0.2, and 0, retaining the same initial mean anomaly of b and all other initial parameters.

In the strongly excited survivor, a circular b still drives d to $e_{d,\max}=0.952$, while the quantity $j_{z,d}=\sqrt{1-e_d^2}\cos I_{bd}$ remains nearly constant, consistent with quadrupole Kozai--Lidov exchange.  Restoring b's RV-drawn eccentricity increases $e_{d,\max}$ to 0.979 and broadens the inclination variations.  The nominal octupole-strength parameter in the test-particle limit is
\begin{equation}
\epsilon_{\rm oct}=\frac{a_d}{a_b}\frac{e_b}{1-e_b^2}\simeq0.16
\end{equation}
\citep{LithwickNaoz2011}, indicating appreciable octupole modulation.  The stronger variation of $j_{z,d}$ in the eccentric-b controls is consistent with departures from the quadrupole invariant.

The response depends on geometry.  The near-coplanar prograde case maintains $e_{d,\max}=0.419$--0.430 across the three controls.  In contrast, the near-polar case collides with the star after 54.5 kyr at $e_b=0.2$ (Figure~\ref{fig:longterm}a), while its circular-b and RV-eccentricity counterparts survive to 1 Myr.  Changing $e_b$ alters the secular frequencies and phases as well as the excitation amplitude, so finite-time survival need not improve monotonically as $e_b$ decreases.  These controls show that b's high eccentricity can amplify secular excitation, while high-inclination Kozai--Lidov exchange alone can already produce extreme eccentricities.

\subsection{Long-term evolution}\label{sec:longterm}

\begin{figure*}[t!]
\centering
\includegraphics[width=\textwidth]{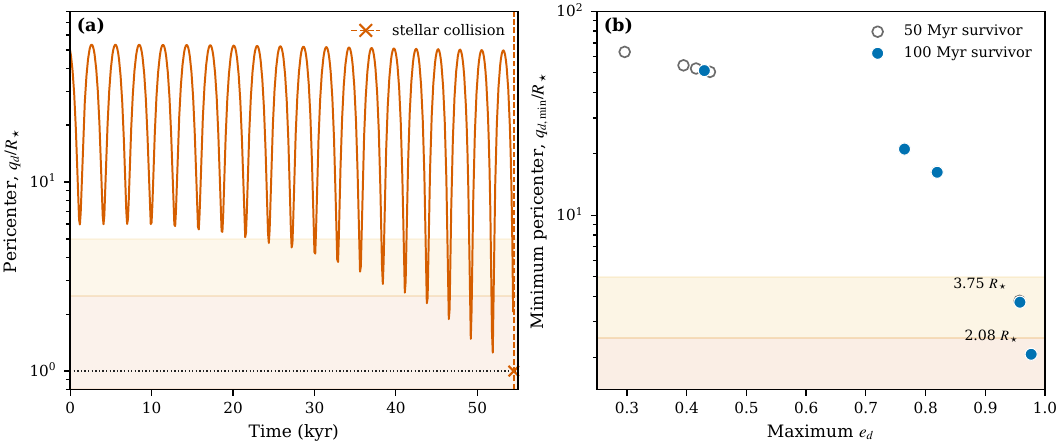}
\caption{Pericenter evolution and long-term outcomes.  Panel (a) follows a near-polar b+d control with $e_b=0.2$: secular excitation drives d into the star after 54.5 kyr.  The dotted horizontal line marks the stellar surface.  Panel (b) gives $q_{d,\min}/R_\star$ versus $e_{d,\max}$ for the nine 50 Myr survivors and the five representatives that reach 100 Myr.  The shaded bands identify passages within 2.5 and 5 stellar radii; Table~\ref{tab:survivors} lists the five 100 Myr cases.\label{fig:longterm}}
\end{figure*}

\begin{deluxetable*}{lcrrrrr}
\tablenum{6}
\tablecaption{Representative b+d Survivors at 100 Myr\label{tab:survivors}}
\tabletypesize{\scriptsize}
\tablewidth{0pt}
\tablehead{\colhead{Case} & \colhead{Geometry} & \colhead{$I_{bd,0}$ ($^\circ$)} & \colhead{$M_d/M_\oplus$} & \colhead{$e_{d,0}$} & \colhead{$e_{d,\max}$} & \colhead{$q_{d,\min}/R_\star$}}
\startdata
00397 & retrograde endpoint & 168.29 & 17.26 & 0.340 & 0.765 & 21.11 \\
00712 & retrograde endpoint & 152.23 & 14.76 & 0.331 & 0.819 & 16.25 \\
00755 & retrograde transition & 108.13 & 16.88 & 0.363 & 0.977 & 2.08 \\
00935 & prograde transition & 72.77 & 15.28 & 0.425 & 0.958 & 3.75 \\
00849 & prograde endpoint & 3.95 & 20.52 & 0.417 & 0.430 & 51.20
\enddata
\tablecomments{The same cases are shown in Figure~\ref{fig:longterm}b.  All rows completed the 100 Myr Newtonian b+d integration.  The two transition cases with $q_{d,\min}<4R_\star$ are strongly excited point-mass survivors.}
\end{deluxetable*}

Delayed stellar collisions remain common after the one-Myr screen.  Among the 39 systems selected across the mutual-inclination bins, 27 collide with the star between 1.03 and 9.36 Myr, leaving 12 at 10 Myr.  Three of those 12 collide at 11.66, 13.04, and 18.90 Myr, and nine complete 50 Myr.  Five of the six representative configurations complete 100 Myr; one collides with the star at 67.70 Myr.  The five survivors cover a prograde endpoint, both prograde and retrograde transition geometries, and two retrograde endpoints.

In a near-polar control, the eccentricity rises from 0.45 to above 0.98 while the pericenter descends to the stellar surface, producing a collision after 54.5 kyr.  Two 100 Myr survivors reach $q_{d,\min}=0.0207$ and 0.0115 au, equal to 3.75 and $2.08R_\star$.  These cases enter a regime where tides, rotational distortion, and the Roche geometry can modify the secular cycle \mbox{\citep{FabryckyTremaine2007,LiuEtAl2015}}.  The other three maintain $q_{d,\min}/R_\star=16.25$, 21.11, and 51.20.

Figure~\ref{fig:longterm}b compares the maximum eccentricity and minimum pericenter attained during the integrations by the 50 and 100 Myr survivors.  Two of the five 100 Myr survivors reach within $4R_\star$, whereas the other three remain beyond $16R_\star$ throughout.

We use conservation of the total Newtonian energy to assess numerical integration error.  Among the 100 Myr runs, the largest initial-to-final relative energy change is $1.13\times10^{-12}$.  Repeating a strongly excited survivor with $\epsilon=10^{-11}$ changes $e_{d,\max}$ by $1.9\times10^{-4}$ and $q_{d,\min}$ by $9.0\times10^{-5}$ au.  The default ($\epsilon=10^{-9}$) and tighter integrations also recover a stellar collision for the same failing case, at 67.697 and 14.631 Myr, respectively.  Both accuracy settings therefore reproduce the stellar-collision outcome, whereas the event time is not converged, consistent with chaotic divergence of the secular phase \citep{HussainTamayo2020}.

\section{Discussion}\label{sec:discussion}

\subsection{The dynamically viable architecture}\label{sec:architecture}

Figure~\ref{fig:longterm} and Table~\ref{tab:survivors} summarize the long-term outcomes.  The dominant constraint on the eccentric solution is the three-dimensional b--d geometry.  Prograde and retrograde configurations near the coplanar endpoints keep the pericenter of d comparatively well separated from the star, giving the U-shaped survival curve in Figure~\ref{fig:ushape}.  Mutual inclinations near $90^\circ$ enable large Kozai--Lidov oscillations, and the eccentric orbit of b adds strong octupole modulation.  Most sampled trajectories in this region end in a stellar collision.  The long-term survivors show that transition geometries can persist for 100 Myr, although their secular cycles can reach pericenters of a few stellar radii.

The combined results favor an architecture in which b and d occupy either a prograde or retrograde near-coplanar endpoint.  A smaller transition region remains dynamically accessible.  Additional constraints on the sky-plane inclination and node of d would distinguish these geometries directly.  The secular route from $e_{d,0}\leq0.1$ shows that the current eccentricity may reflect the phase of a long-term cycle and can be generated after the planets reach their present semimajor axes.

The matched Beta-prior test changes the eccentricity posterior only modestly, while the independently widened phase and apsidal priors retain a solution near $e_d=0.4$.  Improved phase coverage offers the clearest way to determine which part of the allowed range is preferred by the data.

\subsection{Relation to previous dynamical limits}\label{sec:comparison}

\citet{HatzesEtAl2022} drew orbital solutions from their three-planet fit posteriors, sampled $i_d$ between $20^\circ$ and $90^\circ$, and performed preliminary 20 Myr integrations.  Within their RV solution and sampled orientations, stability favored $e_d<0.3$, $M_d<20\,M_\oplus$, and $i_d>40^\circ$.

Our integrations use RV distributions centered at higher eccentricity and sample an isotropic distribution of d's orbital pole, including both prograde and retrograde configurations.  The one-Myr b+d survivors include initial eccentricities up to $e_{d,0}=0.515$ and true masses up to $75.7\,M_\oplus$ (Section~\ref{sec:orientation}; Figure~\ref{fig:ushape}c).  Longer integrations reveal substantial delayed instability: 27 of the 39 systems selected across the mutual-inclination bins collide with the star before 10 Myr, while five representative b+d systems survive to 100 Myr (Section~\ref{sec:longterm}).  These results show that short-term survival admits configurations that can subsequently be removed by longer integrations, alongside long-lived realizations of the eccentric solution.

Preserving the inner planet provides an additional constraint.  In the matched one-Myr controls, adding c to b+d+GR reduces the surviving configurations from 15 to six or seven, depending on the adopted node of c; c itself is frequently the first planet lost (Section~\ref{sec:cgr}).  The earlier 20 Myr limits and our one-Myr b+d survivors therefore describe different integration durations, physical models, and sampled orbital distributions.  Across our experiments, b--d mutual inclination provides the leading geometric constraint, b drives d's secular eccentricity evolution, and the requirement to preserve c further restricts the viable full-system architecture.

\subsection{Observational prospects and model scope}\label{sec:future}

Additional high-precision RV measurements spanning complete d cycles, with dense coverage near the predicted pericenter phases, can reduce the covariance among $e_d$, $\omega_d$, and $t_{0,d}$.  A joint treatment of correlated stellar and instrumental noise would provide a complementary test of the eccentricity posterior.  Astrometric constraints on d would be especially valuable because the stability map is primarily a function of mutual inclination.

The representative integrations do not estimate a full-system survival fraction over the stellar age of approximately 4 Gyr \citep{HuberEtAl2022}.  Extending that inference would require a larger b+c+d+GR posterior sample, coverage of the unknown node of c, and longer integrations.  Trajectories reaching a few stellar radii would also require equilibrium tides, rotational and tidal precession, and Roche-limit checks to distinguish tidally modified systems from disrupted ones.  These effects chiefly concern the most strongly excited transition cases and do not alter the present identification of b as the driver of d's secular evolution.

\section{Summary}\label{sec:summary}

We reconstructed the orbit of the candidate \pimen\ d from 772 published RV measurements and propagated correlated RV posterior samples into three-dimensional integrations.  The main results are:
\begin{enumerate}
\item The RV solution remains near $e_d=0.4$ under wider phase and apsidal priors.  Exact reweighting to a population-informed Beta eccentricity prior gives $e_d=0.378^{+0.050}_{-0.049}$.
\item The b--d mutual inclination is the leading dynamical constraint.  In the 20 kyr screen, 84.5\% of configurations near the prograde and retrograde endpoints survive, compared with 21.3\% across $60^\circ\leq I_{bd}<120^\circ$.
\item Secular forcing by the massive eccentric planet b can move initially near-circular d orbits into $0.3\leq e_d\leq0.5$.  Kozai--Lidov exchange and octupole modulation account for the inclination dependence and the stellar-collision channel.
\item Five representative b+d systems complete 100 Myr integrations.  Two transition cases approach within four stellar radii and are classified as strongly excited point-mass survivors.
\item Matched one-Myr b+c+d+GR integrations retain fewer systems than the b+d model.  Many additional failures involve c becoming unstable first, so preserving the inner planet further restricts the viable three-dimensional architecture.  Among common survivors, c produces small typical changes in $e_{d,\max}$, while GR competes with b-driven precession for c.
\end{enumerate}

The high-eccentricity RV solution has dynamically viable realizations.  The outer giant can excite d into the observed eccentricity range while constraining its viable orbit toward the prograde and retrograde near-coplanar endpoints.  A smaller set of secularly active transition configurations also survives, although short-range physics may modify those reaching the smallest pericenters.

\section*{Data and Code Availability}

The machine-readable RV data, posterior weights, numerical initial conditions, outcome tables, integration protocols, and analysis code underlying this article are available in a versioned Zenodo repository at \url{https://doi.org/10.5281/zenodo.22540472}.

\begin{acknowledgments}
The author thanks the anonymous referee for a careful and constructive review that improved the clarity and rigor of the manuscript.  
This research has made use of the Astrophysics Data System, funded by NASA under Cooperative Agreement 80NSSC21M0056.  This research has made use of the VizieR catalogue access tool, CDS, Strasbourg, France (DOI: 10.26093/cds/vizier; \citealt{OchsenbeinEtAl2000}).  Generative-AI tools were used to assist with code implementation and language editing.  The scientific analysis, interpretation, and conclusions were independently validated by the author, who takes full responsibility for the content of this work.
\end{acknowledgments}

\software{REBOUND v4.6.0 \citep{ReinLiu2012}, REBOUNDx v4.6.2 \citep{TamayoEtAl2020}, RadVel v1.6.1 \citep{FultonEtAl2018}, dynesty v3.1.0 \citep{Speagle2020}, NumPy v2.2.6 \citep{HarrisEtAl2020}, SciPy v1.15.3 \citep{VirtanenEtAl2020}, pandas v2.3.3, Matplotlib v3.10.7 \citep{Hunter2007}}

\bibliographystyle{aasjournalv7}
\bibliography{references_observational,references_dynamics}

\end{document}